# 2D Materials for Future Heterogeneous Electronics

Max C. Lemme[1,2,*], Deji Akinwande[3], Cedric Huyghebaert[4], Christoph Stampfer[5,6]

[1] Chair of Electronic Devices, RWTH Aachen University, Otto-Blumenthal-Str. 2, 52074 Aachen, Germany

[2] AMO GmbH, Advanced Microelectronic Center Aachen, Otto-Blumenthal-Str. 25, 52074 Aachen, Germany

[3] Department of Electrical and Computer Engineering, Microelectronics Research Center, The University of Texas at Austin, Austin, 78712, Texas, USA

[4] IMEC, Leuven, Belgium

[5] JARA-FIT and 2nd Institute of Physics, RWTH Aachen University, 52074 Aachen, Germany, EU

[6] Peter Grünberg Institute (PGI-9), Forschungszentrum Jülich, 52425 Jülich, Germany, EU

**Abstract**

Graphene and two-dimensional materials (2DM) remain an active field of research in science and engineering over 15 years after the first reports of 2DM. The vast amount of available data and the high performance of device demonstrators leave little doubt about the potential of 2DM for applications in electronics, photonics and sensing. So where are the integrated chips and enabled products? We try to answer this by summarizing the main challenges and opportunities that have thus far prevented 2DM applications.

**Manufacturing Technology**

The key answer to this question, in our opinion, can be found by comparing the manufacturing readiness level of 2DM with standard semiconductor technology. What is needed, but not available yet, are turnkey manufacturing solutions that bring 2DM into silicon (Si) semiconductor factories. These "unit processes" then serve to integrate 2DMs with Si complementary metal oxide semiconductor (CMOS) chips in the back-end or front-end of the line.[1,2] Deposition and growth technology of 2DMs is generally available at the wafer scale, but defects and contaminations are not yet compliant with specifications defined for production.[3] In addition, high process temperatures are typically required for high quality materials, which complicates direct growth on wafers and makes transfer technologies desirable. In principle, wafer bonding techniques could solve this, but have not reached full manufacturing levels.[4] At the device level, challenges are linked to the control of the dielectric- and contact interfaces to the 2DMs. The self-passivated nature of 2DM surfaces requires seeding to achieve the deposition of dielectrics with manufacturable methods, e.g. through atomic layer deposition. The resulting non-ideal interfaces limit device performance compared to the best laboratory demonstrators that use crystalline 2D insulators such as hexagonal boron nitride.[5] The same is true for electrical contacts to 2DMs, which only partly meet industry specifications,[6] and have not reached manufacturing readiness. The removal or etching of materials with high selectivity towards underlying layers is particularly challenging for 2DMs, because it requires atomic precision that can only be achieved with specific chemistry and dedicated atomic layer etching tools. Developing suitable processes will be tedious, because of the wide range of potential 2DMs and their combinations. As a result, etching chemistry and other, physical process parameters depend strongly on the specific situation which each require individual solutions. Doping, or the replacement of atoms in the crystal lattice, is a standard but crucial technology for silicon that relies on statistical distributions. In the 2DM field, the term doping is often used to describe charge transfer to the 2D layer from defects or molecular adsorbates in its vicinity. Controlling this "effective-doping" with precision and long-term stability remains a challenge, but so would classic doping, which would ideally require replacing 2D crystal atoms in a deterministic way, as shown for silicon technology.[7] Solving these crucial manufacturing bottlenecks is the explicit goal of the European Experimental Pilot Line for 2D Materials.[8] Co-integration of 2DMs with silicon CMOS technology will lead to a vast increase in chip functionality and enable the arrival of 2DM applications in the order of their device complexity, as illustrated in Figure 1 and presented in the following.

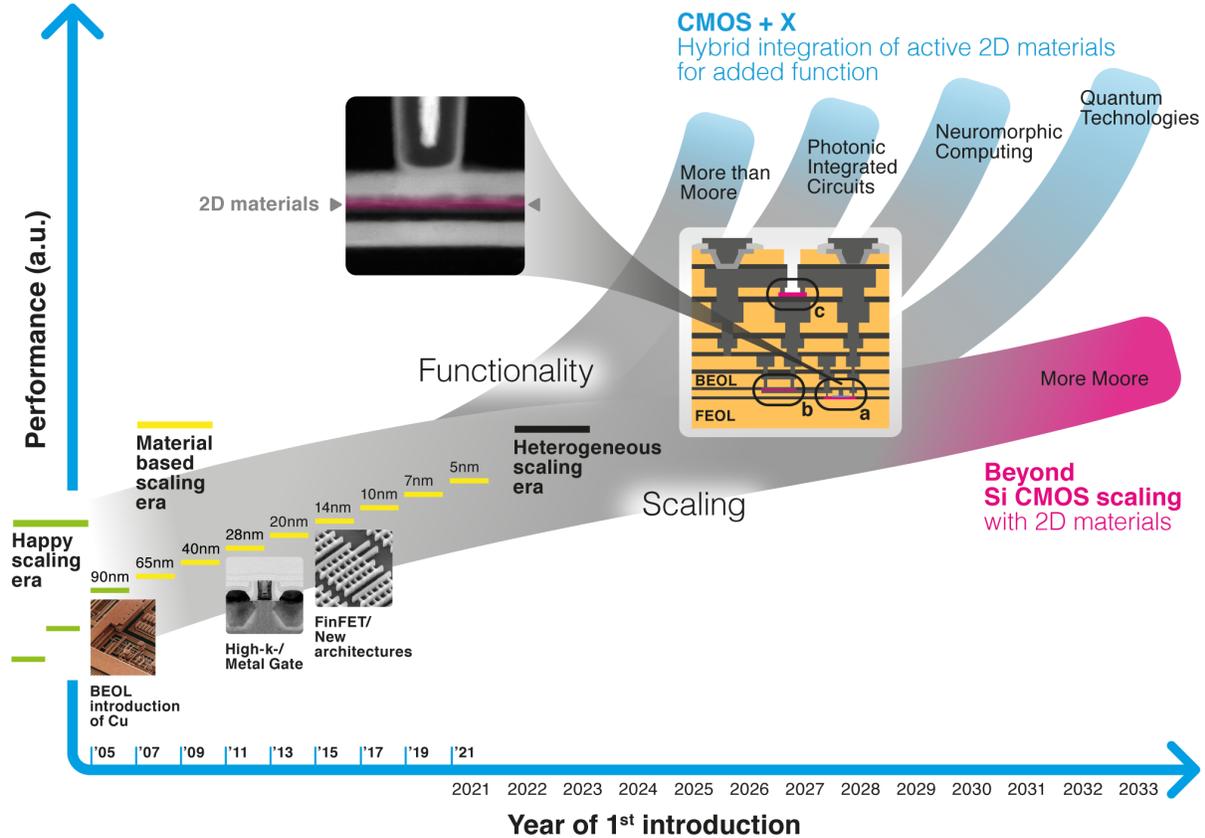

Figure 1: The era of geometrical or Dennard[9] scaling of silicon technology ended around the turn of the century (green lines, "happy scaling"). Since then, material and architecture innovations like copper interconnects,[10] high-k dielectrics with metal gates[11] and FinFETs[12] continued to drive Moore's law (yellow lines, "less happy scaling"). Future scaling, or "More Moore", may require thin nanosheet transistors, where 2D materials are considered ideal candidates (magenta, inset a and transmission electron micrograph).[13,14] Substantial performance and functionality gains are expected through "CMOS + X" integration, for example through sensors or high frequency electronics integrated on CMOS chips in the "More Than Moore" domain (inset c). Photonic integrated circuits may boost overall system performance and data handling capabilities, as well as unlock spectroscopic sensing applications, enabled by the optoelectronic performance of 2DMs. Computing-In-Memory or memristors will enable future

neuromorphic computing applications and 2DMs may be ideally suited to be integrated with silicon CMOS (inset b). 2D quantum technologies are the least mature even at the laboratory level, but will benefit from all expected achievements as 2DMs enter semiconductor processing lines. 2D materials hold great promise to become the X-Factor for CMOS. This may be described as the era of heterogenous scaling, where new and additional materials provide unprecedented performance in three-dimensional chip stacks. Note that the Y-axis had a unit of "log2(#transistors/$)" during the classic "Moore's law" period. This has to be replaced in the era of heterogeneous scaling, and we suggest labeling it „Performance (a.u.)", because the increase in performance will become application specific. It will be determined by (combined) factors like power consumption and efficiency, capability to perform pattern recognition, sensor fusion etc., which results in somewhat arbitrary units due to the diverse functionalities and underlying technologies. (Insets: BEOL Introduction of Cu: Reproduced with permission from the AAAS, reference[10]; High-k/Metal Gate: © 2007 IEEE. Reprinted, with permission, from Mistry, K. *et al.* A 45nm Logic Technology with High-k + Metal Gate Transistors, Strained Silicon, 9 Cu Interconnect Layers, 193nm Dry Patterning, and 100% Pb-free Packaging. in *2007 IEEE International Electron Devices Meeting* 247–250 (2007)[11]; FinFET / New architectures: Republished with permission of IEEE, from Jan, C.-H. et al. A 22nm SoC platform technology featuring 3-D tri-gate and high-k/metal gate, optimized for ultra low power, high performance and high density SoC applications. in 2012 International Electron Devices Meeting 3.1.1-3.1.4, 2012[12]; permission conveyed through Copyright Clearance Center, Inc.; 2D Materials: source: reference [14]).

**More Moore**

In general, gains in advanced semiconductor technology nodes are enabled through increased complexity of the integration architectures as well as holistic system-technology co-optimization. On the device level, leading semiconductor manufacturers are turning from FinFETs to stacked nanosheet FET architectures for the most advanced nodes of CMOS technology.[15] These nanosheet devices are currently still based on Si channels. Different flavors of such nanosheet architectures are in evaluation for future technology nodes, e.g. a so-called fork sheet design that allows tighter n-to-p spacing,[16] or integration of p and n-type nanosheets on top of each other.[17] Further scaling of the channel length requires shrinking the channel thickness by a similar factor to guarantee sufficient electrostatic control to suppress short channel effects. Trimming down the Si sheet thickness to the required values increases the charge scattering at the interfaces and results in a dramatic drop of the carrier mobility in the channel,[18] which ruins device performance. 2D semiconductors would be the ultimate version of nanosheets, because they are self-passivated in the 3rd dimension and charge carrier mobility is not strongly affected from surface scattering. Therefore, mobility remains high even at the atomic thickness limit. This unique behavior in principle enables real scaling for a several technology nodes and is a strong incentive for the semiconductor industry to finally consider replacing silicon as the transistor channel material for future advanced nodes.[13,14,19] This takes us back to the fundamental technical and scientific challenges which are linked to 2D integration. Here, identifying a suitable gate oxide stack and finding low ohmic contact schemes are particularly important. The former is essential to preserve the 2D material properties and to provide sufficient electrostatic control while limiting gate leakage currents.[5] 2D hexagonal boron (hBN) nitride has been widely applied to demonstrate high performance devices based on 2DMs, but its band gap and band offsets dictate that sufficient electrostatic control can only be achieved with one or two monolayers. This additional boundary condition leads to intolerably high gate leakage currents and other solutions will have to be found.[20] Low ohmic contacts are required to maintain the benefits of the channel material in integrated circuits, because high resistance contacts can dominate and severely limit the integrated device performance.[21] Recently, substantially improved electrical contact resistance to $MoS_2$ has been reported by using semi-metallic Bismuth, which strongly suppresses metal induced gap states and the spontaneous formation of degenerate states in the $MoS_2$.[22] Nevertheless, more breakthroughs like this are needed to uncover and fully exploit the potential of monolayer transistors in CMOS circuits, to revive transistor downscaling and continue Moore's law.

**More Than Moore**

Applications in this category are likely first to enter the market, because they are manyfold, yet often very specific, so that they may tolerate defects and larger device variations.

2DMs are well-suited for gas, chemical and biosensing, because of their inherently high surface to volume ratio and versatile functionalization.[23] Thus, any charged particle or molecule in the vicinity of certain 2D layered materials can modify their conductivity. However, ideal 2DMs are chemically inert, which means that chemically active defect sites strongly enhance the reactivity of 2D based sensors. Precise defect engineering is therefore essential for controlling sensitivity. In addition, sensor selectivity is essential. It may be achieved through surface functionalization or via arrays made from different sensors to mimic complex biological systems like the nose. Here, the portfolio of 2DMs with diverse sensor "fingerprints" may be utilized in conjunction with machine learning algorithms for sensor read-out.

Micro- and nanoelectromechanical systems (MEMS/NEMS) typically rely on mechanically movable parts on the chip. 2DMs exhibit exceptional mechanical properties that produce ultra-thin membranes, which translates directly to extremely high sensitivities in piezoresistive and opto-mechanical read out schemes, providing efficient signal transduction in NEMS. 2D membrane-based NEMS applications include pressure sensors, accelerometers, oscillators, resonant mass sensors, gas sensors, Hall effect sensors, and bolometers.[24]

2DMs possess a range of advantages over existing technologies for optoelectronic and photonic applications,[25] in particular outside of the spectral range that can be addressed with silicon. But even there, the direct band gaps of many semiconducting 2DMs provide advantages over silicon when it comes to light emission.[26] Semi-metallic and small band gap materials like graphene, platinum diselenide or black phosphorous open up the infrared (IR) regime, where they compete with often costly III-V semiconductor technologies. Although the 2D nature translates into low absolute absorption in the vertical direction, the combination with IR-sensitive absorption layers leads to high detector responsivity.[27]

**Photonic Integrated Circuits**

Photonic integrated circuits are considered as ultimate performance boosters for data transmission on and across computer chips.[28] Connecting them to silicon electronics through optoelectronic converters at extremely high data rates is a key enabling technology. 2DMs, and in particular graphene, can be transferred onto photonic waveguides and provide wide-band photodetection and modulation.[25,29,30] By removing the need for epitaxy, 2D-based photonics thus allows integrating active device components with Si photonics, but also with passive amorphous waveguide materials, like silicon nitride. This opens the door for facile integration of complex photonics applications on top of CMOS. In fact, some 2DMs like platinum diselenide can also be directly and conformally grown at temperatures below 400°C,[31] which is a clear advantage in the quest to co-integrate photonic integrated circuits with silicon CMOS technology.[32] With the potential for integrated 2D light sources,[26] 2DMs could ultimately enable the convergence of electronics and photonics and bridge the spectrum across the THz gap.

**Neuromorphic Computing**

Neuromorphic computing aims to provide brain-inspired computing devices and architectures to realize energy efficient hardware for artificial intelligence applications.[33] On the device level, requirements for neuromorphic computing include merging memory with logic to enable Computing-In-Memory and memristive device characteristics that mimic synapses and neurons. The former can already be realized with conventional memory technologies while the latter translates to threshold switches, and non-volatile memristors with a wide range of programmable resistance states.[34,35] Though relatively nascent, 2D memristors have shown promising performance including switching energies on the order of zeta-joules, sub nano-second switching times, dozens of programmable states, and prototype artificial neural networks at the wafer-scale.[36] This may enable applications in sensor systems and edge computing, for example by preprocessing of sensor data or on-chip sensor fusion.[37] In addition to neuromorphic computing, 2D memristive devices have been shown to provide a wide range of non-computing functions including physically unclonable functions for security systems, and radio-frequency switching for communication systems.[38]

From a scientific view, the phenomena of resistive switching in 2D devices have been attributed to ionic transport,[39,40] defect/filament formation[41] or phase change effects.[42] Notwithstanding these fundamental aspects, 2D memristive switching remains a topic enjoying increasing discussion and research. At the device level, a fundamental challenge is improving the number of times the resistance can be switched, so-called endurance, which requires further studies into the aging effect of the underlying mechanism(s). Similarly, improving material uniformity will be essential in order to realize massively connected device arrays that can mimic the hyper-connectivity and efficiency of the brain. Hearteningly, over a dozen 2DMs have exhibited memristive effect with this collection likely to grow in the coming years. Therefore, computational methods will be increasingly needed to guide experimental studies and optimize memristive devices for maximum performance.

**Quantum Technologies**

2D materials and related van-der-Waals heterostructures offer also a variety of properties that make them highly tunable quantum materials interesting for spintronics and future quantum technologies.[43] Two-dimensional material systems not only have the ability to enable artificial states of quantum matter, but fulfil a number of promises for solid-state quantum computing, to serve as key components in quantum communication circuits or allow interesting quantum sensing schemes. Indeed, 2DMs are a promising solid-state platform for quantum-dot qubits, as recognized quite early,[44] for topological quantum computing elements, as well as coherent sources of single-photon emitters.

Quantum computing based on semiconductor quantum dots (DQs) uses individual spin states of trapped electrons. It relies, for example, on long spin coherence times with the host material. This makes graphene a very interesting materials for spin qubits, thanks to its weak spin-orbit coupling (carbon atoms are very light) and weak hyperfine coupling (Carbon 12 is nuclear spin free). With the recent progress in confining single-electrons in gate-controlled QDs in gapped bilayer graphene,[45,46] first spin-qubits are now around the corner. The possibility to make spin qubits in 2DMs will also allow evaluating the additional valley degree of freedom as possible qubit states; interesting proposals of valley- and spin-valley qubits exist. In addition, stationary qubits in 2DMs may afford coupling to photonic qubits realized by single-photon emitters (SPEs), for example in nearby wide band-gap hexagonal boron nitride or semiconducting transition metal dichalcogenides, such as $WSe_2$.[47] In these 2DMs, SPEs have been demonstrated in recent years and this opens the door to distributed quantum networks, where photonic qubits could act as interlinks that entangle distant stationary qubits, e.g., spin-qubits. Such robust, bright and indistinguishable single-photon emitters are essential for creating photonic (flying) qubits for efficient quantum communication.

2D heterostructures are, moreover, promising materials for topological quantum computing, where quantum states are better (i.e. topologically) protected against disorder, compared to standard quantum computing.[48] Combining, for example, a quantum anomalous Hall insulator or graphene tuned into the canted anti-ferromagnetic quantum Hall phase with s-wave superconductors is a promising platform for topological quantum computing. In short, these advances make 2DMs and their heterostructures in many ways an exciting and promising platform for future quantum technologies.

**Conclusions**

2DMs provide exceptional performance benefits over existing technologies at the device level. They also carry the promise of easy integration with silicon CMOS technology. This makes them prime candidates for substantially expanding the functionality of silicon chips, also referred to as "CMOS + X". We are confident that 2DMs will increasingly become such an X-factor in future integrated products. The bottleneck towards 2D material-based heterogeneous electronics is reaching the required manufacturing readiness levels, which may be different depending on the targeted application.

**Acknowledgements**

The authors acknowledge funding from the European Union's Horizon 2020 research and innovation programme under grant agreements 881603 (Graphene Flagship), 952792 (2D-EPL), 101017186 (AEOLUS), 101016734 (MISEL), 971398 (ULTRAPHO), 101006963 (GreEnergy), 956813 (2Exciting), 863337 (WIPLASH), 825272 (ULISSES) and 829035 (QUEFORMAL) as well as the European Research Council (ERC) under grant agreements 820254 and 307311. We further acknowledge support from the Deutsche Forschungsgemeinschaft (DFG, German Research Foundation) under Germany's Excellence Strategy - Cluster of Excellence Matter and Light for Quantum Computing (ML4Q) EXC 2004/1 -390534769 and through the DFG grants STA 1146/11-1, STA 1146/12-1, LE 2440/7-1, LE 2440/8-1 and LE 2440/11-1. Furthermore, support by the Bundesministerium für Bildung und Forschung (BMBF, German Ministry of Education and Research) through the grants 03XP0210 (GIMMIK), 16ES1121 (NobleNEMS) and 16ME0399 (NEUROTEC II) is acknowledged, as is the Alexander von Humboldt-Foundation. Part of the work leading to this manuscript has been carried out in the Aachen Graphene & 2D Materials Center.

**Author contributions**

MCL, DH, CH and CS co-wrote the manuscript and co-created the figure.

**Competing interests**

There is no competing interest.

**Data availability statement**

Data sharing not applicable.